\documentclass[a4paper,onecolumn,10pt,accepted=2026-03-30]{quantumarticle}
\pdfoutput=1
\usepackage[utf8]{inputenc}
\usepackage[english]{babel}
\usepackage[T1]{fontenc}
\usepackage{amsmath}
\usepackage{hyperref}
\usepackage{graphicx}
\usepackage{tikz}
\usepackage{tikz-cd}
\usepackage{lipsum}
\usepackage{amssymb}
\usepackage{physics}
\usepackage{amsthm}
\usepackage{ytableau}
\usepackage{amsfonts}
\usepackage{diagbox}
\usepackage{multirow}
\usepackage{ulem}

\usepackage[numbers,sort&compress]{natbib}
\usepackage{bm}
\newcommand{\mrk}[1]{{\color{black} #1}}
\newcommand{\ks}[1]{{\color{black} #1}}

\newcommand{\marcin}[1]{{\color{black} #1}}

\begin{document}

\title{``Nonlocality-of-a-single-photon'' based Quantum Key Distribution and Random Number Generation schemes and their  device-independent security analysis}

\author{Konrad Schlichtholz}
\affiliation{International Centre for Theory of Quantum Technologies (ICTQT),
University of Gdańsk, 80-309 Gdańsk, Poland}
\orcid{0000-0001-8094-7373}
\email{konrad.schlichtholz@ug.edu.pl}
\author{Bianka Woloncewicz}
\affiliation{International Centre for Theory of Quantum Technologies (ICTQT),
University of Gdańsk, 80-309 Gdańsk, Poland}
\affiliation{Quantum Research Center, Technology Innovation Institute, Abu Dhabi UAE}
\author{Tamoghna Das}
\affiliation{International Centre for Theory of Quantum Technologies (ICTQT),
University of Gdańsk, 80-309 Gdańsk, Poland}
\affiliation{Department of Physics, Indian Institute of Technology Kharagpur, Kharagpur 721302, India}
\orcid{0000-0002-8074-6720}
\author{Marcin Markiewicz}
\affiliation{International Centre for Theory of Quantum Technologies (ICTQT),
University of Gdańsk, 80-309 Gdańsk, Poland}
\affiliation{Institute of Theoretical and Applied Informatics, Polish Academy of Sciences, ul. Ba{\l}tycka 5, 44-100 Gliwice, Poland}
\orcid{0000-0002-8983-9077}
\author{Marek \.Zukowski}
\affiliation{International Centre for Theory of Quantum Technologies (ICTQT),
University of Gdańsk, 80-309 Gdańsk, Poland}
\orcid{0000-0001-7882-7962}

\maketitle

\begin{abstract}
 The question of ``non-locality of a single photon'', which started with a paper by  Tan, Walls and Collett (TWC, 1991) stirred a thirty years long debate. This hampered attempts to use the TWC interferometric scheme in quantum cryptography. The scheme involves a single photon 50-50 beam-split into two modes propagating to two different spatially separated observation stations at which weak homodyne measurements are made. The physics and non-classicality of such an arrangement has been understood only recently, and points out that an unquestionable Bell non-classicality, as was suggested by Hardy (1994), can be observed when the local measurement settings differ by the weak local oscillator being \textit{on} or \textit{off}, and additionally the homodyning for the \textit{on} case is not balanced. Based on that,
we present a single-photon based device-independent quantum key distribution scheme secure even against (the unphysical) no-signaling eavesdropping. In our protocol the random bits of the cryptographic key are obtained by measurements on the single photon, that is for \textit{off} settings at both Alice and Bob sides, while the security is positively tested if for eavesdropping testing runs one observes a violation of a specific Bell inequality involving the  \textit{on} and \textit{off} weak homodyne measurements as alternative local settings. The security analysis presented here is based on a decomposition of the correlations into extreme points of a no-signaling polytope, which allows for identification of the optimal strategy for any eavesdropping constrained only by the no-signaling principle. For this strategy, the key rate is calculated, which is then connected with the violation of a specific Clauser-Horne inequality. We also adapt this analysis to propose a self-testing quantum random number generator based on the old idea that employs the randomness of reflection and transmission events of a quantum light impinged on a 50-50 beamsplitter.
\end{abstract}

\section{Introduction}

A single photon impinging on a 50-50 beamplitter is intuitively {\it the} quantum method to get random bits. Excitation of two spatially separated optical modes by just one single photon gives an entangled state of the modes.  However, Bell-nonclassical events cannot be  observed in this setup by just using photon-detectors. The trail blazing paper on this is the one of Tan, Walls and Collett \cite{TWC91}, which was aimed at showing that one can obtain a Bell inequality violation via  weak balanced homodyne measurements. This scheme has not been used in quantum cryptography due to the controversies associated with it, first pinpointed  by  \cite{SANTOS}. Since then various  related scenarios  were also proposed, however not involving weak homodyne measurements or the exact single-photon input state, \cite{Hardy94,BAN_WOD,GARYqed,ACINgroup,CASPAR,MUNRO,Enk05,Steering_SP}.

Recently, in papers \cite{modelNJP,SinglephotonNJP,3rdPaper} a comprehensive analysis of the problem of revealing Bell non-classicality of single-photon using weak homodyne measurement was conducted. In particular, the Tan-Walls-Collett setup was finally excluded as possible generator of non-classicality for quantum cryptography by a derivation of an exact local realistic model of the correlations in \cite{modelNJP}. However, the authors propose a modification of the setup that allows for proper violation of a Bell inequality.  As this development allows one to finally think about quantum informational applications of such processes, we shall present here after further modifications to the setup two such applications. Note that this development inspired the formulation of a ``Gisin's Theorem'' for arbitrary entangled states of quantum optical fields, and all-optical measurement devices. That is, devices involving passive optical devices, possibly optical ancilla fields, and (possibly photon-number resolving) photo-detectors \cite{schlichtholz2023generalization}.

One of the most important quantum information tasks is the quantum key distribution (QKD). The perfect form of quantum key distribution involves device independent protocols, which are usually based on a violation of a related Bell inequality. This warrants that the protocol cannot be simulated by a classical, in principle deterministic, model. Importantly, one can treat device-independent formulation of a certain quantum key distribution protocol, as a proof of its inherent quantumness. From the point of view of less than perfect practical realizations, one moves to  semi device independence, or even protocols in which one must trust that the involved devices work according to quantum laws. Still, one can formulate the following rule: {\it if one is not able to formulate a device independent version of a certain supposedly quantum key distribution protocol, then it contains loopholes allowing eavesdropping.} Therefore, our aim here is to show that  generalized TWC correlations involving a beam-split single photon, and unbalanced \textit{on-off} weak homodyne measurements can be used to formulate  a device-independent quantum key distribution protocol. This seems to be a theoretical task, as unfortunately the related Bell-CH inequalities are not violated maximally (see further) and this imposes a high threshold on the transmission efficiency. Nevertheless, there is an important practical element of the considerations: the protocol can be modified and then applied to get a device-independent quantum random number generator.

The great advantage of device-independent protocols is that intrinsic randomness of the generated bit strings does not depend on trusted devices, but relies solely on properties of the outcomes statistics; this however requires Bell non-classicality in the system. The security issues and thresholds of such schemes, in the  presence of a ``quantum'' adversary, see e.g. \cite{Ekert1991,Brunner13,Mayers-Yao,acin-2007-98, Masanes2011, lit12}, or  in the presence of a stronger  ``no-signalling'' (unphysical) eavesdropper, see e.g.  \cite{Kent,AcinGM-bellqkd, masanes-2006, acin-2006-8, hanggi-2009} are discussed in many papers. 

In this paper, we propose modifications to the setup presented in \cite{SinglephotonNJP,3rdPaper}.  We show that a single photon together with weak homodyne measurements enables generation of random string of bits which can be  further used in device-independent QKD,  secure against individual attacks of no-signalling eavesdropper. 
Our modifications, which include the introduction of more experimentally friendly threshold detectors, instead of photon-number-resolving ones, and using stronger local oscillators (but still weak) allow us to obtain an order of magnitude higher violation of Bell inequality than the one presented in \cite{SinglephotonNJP,3rdPaper}. Thus, the modified process and protocol lead to higher secure bit rates in quantum informational applications.  In addition, our proposal exhibits significant simplicity. Namely, we implement the entire QKD protocol within the two-setting-two-outcome Bell-type scenario, which allows  for  description of eavesdropping in terms of the simplest no-signaling polytope. 
Moreover, let us notice that our  scheme uses seemingly complementary features of two emblematic protocols: BB84 \cite{BB84} and Ekert91 \cite{Ekert1991}. The secret key is distributed by a  single photon like in  BB84, while the security of the protocol is based on a Bell test as in device-independent versions of Ekert91.

\section{Experimental Setup}

\begin{figure}\centering
\includegraphics[width= 0.7\columnwidth]{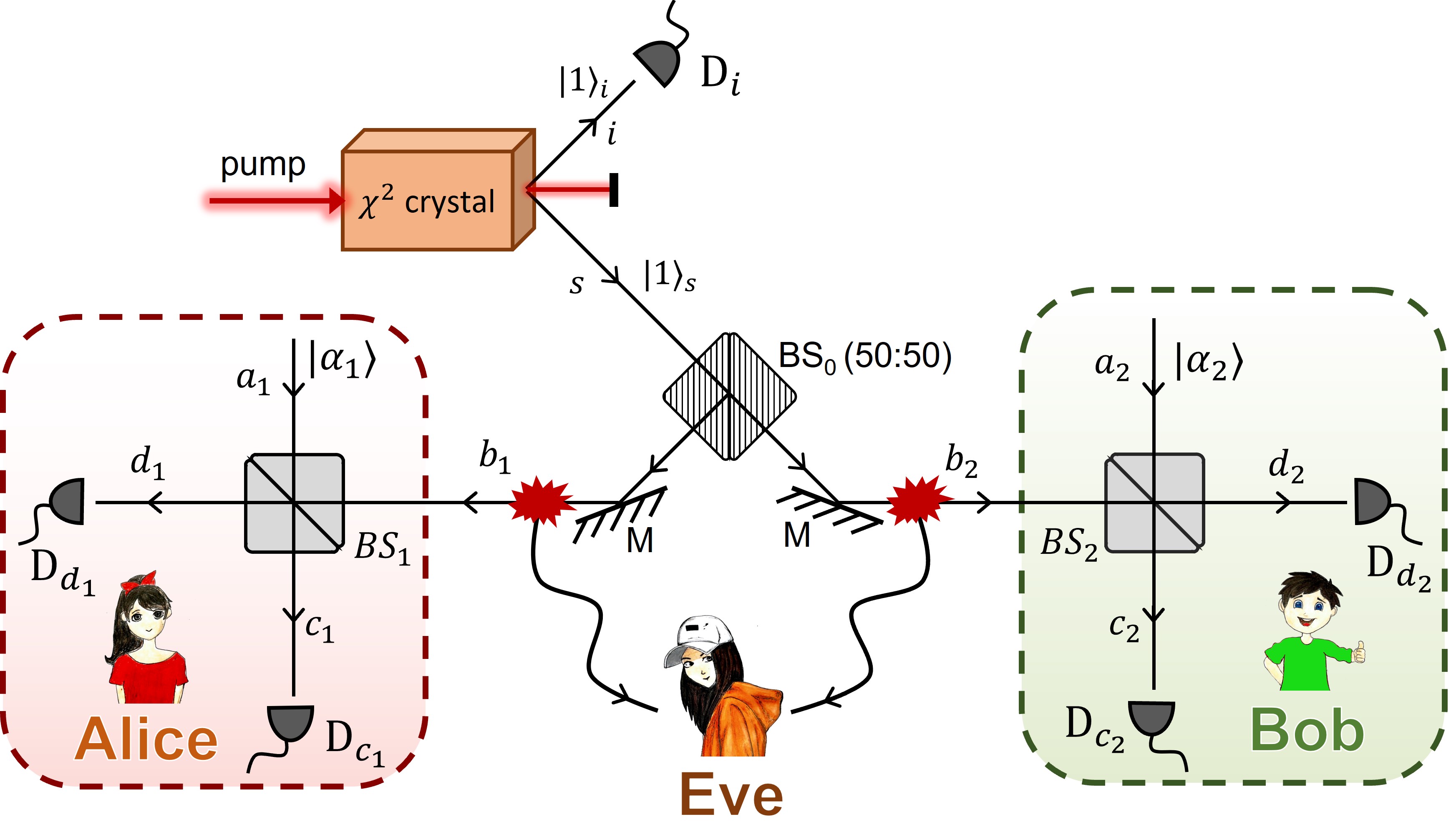}
\caption{\label{schematic}
Schematic diagram of an all-optical setup for quantum key distribution protocol 
\mrk{involving a superposition of single photon excitations of two optical modes, and local on/off weak homodyne measurements in separated measurement stations. A pulsed SPDC source produces pairs of photons. A run of the experiment/protocol begins with a heralding event at detector $D_i$ --- detection of a single idler photon. A single signal  photon is therefore in optical mode $s$. It impinges on a balanced $(50:50)$ beamsplitter $BS_0$. The outputs of it are fed via modes $b_1$ and $b_2$ into two spatially separated  detection stations operated by two observers ~``Alice'' and  ``Bob''}. 
Each of the measurement stations is equipped  with  a beamsplitter $(BS_j)$ with tuneable transmissivity $T_j$. \mrk{Two phase-locked local oscillator coherent pulses $\ket{\alpha}_j$ are also distributed to the stations, in a temporal synchronization with the optical excitations produced by the signal photon}. Both stations implement \mrk{weak unbalanced} homodyne measurements,
\mrk{with measurement settings controlled by  Alice and Bob by adjusting $T_j$ and both the amplitudes and the phases of  $\alpha_j$'s.  They collect data on the local detection events. The quantum key distribution protocol which we present here involves local on/off settings for homodyning, that is $\alpha_j(\textrm{off}) =0$.} }
\end{figure}


\mrk{The scheme of the  interferometric  setup  to be used in the protocols which we propose is presented in the FIG. \ref{schematic}.
Pulse pumped spontaneous parametric down conversion process produces pairs of photons, customary called  signal and idler.
With a  sufficiently weak pump, the state produced in the process is practically 
a superposition of vacuum and a photon pair.  This is due to the fact that the output of a PDC process gives rise to the following superposition of Fock states of the signal and idler modes $|PDC\rangle=\cosh(\Gamma)^{-2}\sum_{n=0}^\infty \tanh^n(\Gamma)|n_{s}, n_{i}\rangle$ where $n_s,n_i$ represent respectively $n$ photons in the signal and idler mode, and $\Gamma$ represents the pumping strength. The relative probability  of having  the two signal-idler pairs compared to just single pair is $\tanh^2(\Gamma)$, which can be made arbitrarily small.

Our aim is to have single photons on which the protocol will be based.
To this end detector $D_i$ is placed in  the idler mode $i$. 
 If it is photon-number resolving and detects one idler photon,  it heralds a signal photon in mode $s$, and the start of a protocol run. If it is a threshold detector, the impact of higher photon events can be  suppressed with the use of a very low pump intensity, discussed above. Note that, one can also use a different type of single photon source, e.g. deterministic ones based on quantum dots, which do not exhibit higher photon number emissions \cite{Huber2017}. In such a case heralding would be unnecessary.
 
 Further on, only runs with detected idler photon  are taken into consideration. This detection marks a successful preparation of the initial state, which can be publicly announced. This information is kept for the classical post-processing round of the protocol during which  the both honest protocol parties keep the measurement outcomes for only those runs in the case of  which the initial state is heralded by the firing of $D_i$. \marcin{Importantly, this communication does not reveal any information about the key as after the detection in the idler mode future outcomes of measurements on the (beam-split) signal mode are uncorrelated with the outcome registered by the detector $D_i$}. The heralding process only guarantees the presence of the single photon in the signal mode $s$. Its presence further on in the output modes of the central beamsplitter, $b_1$ or $b_2$, is in  a 50-50 quantum superposition.  Heralding strategies of this kind are standard in loophole-free Bell tests, and in device independent QKD based on Bell non-classicality \cite{Ent-swapp, Hensen2015,Zhang2022}.  Note that 
 this is a {\em pre-selection} process not a post-selection one. Only the runs of the experiment with single idler detected at $D_i$ are the protocol runs. Information on whatever is wherever registered in runs without this heralding/initiating event is discarded.
 
The heralded in this way single signal photon from the signal mode $s$ impinges on the balanced beamsplitter $BS_0$.
The resulting output field  state is distributed into  two spatially separated measurement stations of Alice and Bob.  The state of the output modes of the beamsplitter $BS_0$ which propagates to the measuring station of  Alice and Bob reads:
\begin{equation}\label{input-state}
    \ket{\psi}_{b_1,b_2}=\frac{1}{\sqrt{2}}(\hat b_1^\dagger+\hat b_2^\dagger)\ket{\Omega}_{b_1,b_2}=\frac{1}{\sqrt{2}}(\ket{01}_{b_1,b_2}+\ket{10}_{b_1,b_2}),
\end{equation}
where $\ket{\Omega}$ is the vacuum state, and $\hat b_j^\dagger$,  denotes the creation operator in the  mode $b_j$, for $j = 1,2$. The modes $b_1$ and $b_2$ feed the respective local measurement stations of Alice and Bob.
The two observers perform local weak homodyne measurements, the settings of which are randomly  chosen, and defined by tuning the transmissivity  $T_j$ of their local beamsplitters $BS_j$ and the amplitude $\alpha_j$ of local oscillators fields in the modes $a_j$. In the case of the latter parameter,  the observers can control the energy of the coherent pulses, and their phases. \marcin{The local protocol settings of the energy are either $|\alpha_j|^2=0$  or a specific non-zero value $|\alpha_j|^2=\alpha^2$, which is optimal for the test of non-classicality.}  The unitary transformation related to $BS_j$ transforms the input modes $a_j$  and $b_j$ into the output modes $c_j$ and $d_j$ which are monitored by the local end detectors. We stress that any type of detector that is able to efficiently enough detect a single photon can be used in this experiment.}


 \mrk{To get the necessary phase stability of the coherent pulses which are used in homodyne measurements at the stations of Alice and Bob, as well overall temporal synchronization in the runs of the experiment, one can use the following procedure. The pulsed pump in Fig. 1 feeding the PDC crystal can be obtained via a frequency doubling procedure of pulses of a laser that feeds the entire interferometric device (not in the picture). The PDC down conversion process should be then frequency degenerate, to the central frequency of the original laser pulses. These laser pulses via a proper beam-splitting sequence, before the frequency doubler, can be used to produce also a pair of time synchronized and phase locked pulses of coherent light, which are also {\em time synchronized with the trigger/heralding events}, if the latter ones  happen. } 
These coherent pulses are distributed to the parties alongside the encoding modes $b_1$ and $b_2$, \mrk{to serve as local oscillators in homodyne measurements}. The distribution could be done for example using double-core optical fibers \cite{Phase_stab}. Such a solution allows for keeping the phase locking between the beams at longer distances. This is because while the phase could \mrk{drift} during the transmission, both modes undergo analogous phase changes in such a fiber.
 The distribution of the coherent reference pulse from the same source  does not cause any security problems, and our  further analysis is entirely device-independent.
\mrk{For suitably space separated measurement stations of Alice and Bob} the locality of the settings choice  \mrk{can be satisfied} in the setup as the settings  \mrk{are} determined by local phase transformations applied to the reference pulses \mrk{and  by the local shutters}.  

In the context of the long-range key distribution to circumvent the problem of losses, one could also consider the construction of the quantum repeater based coherence swapping \cite{coh-swapp} (see also Appendix \ref{app:swaping})

\section{Cryptographic Protocol}

Our cryptographic scheme is based on a Bell-type scenario. The local oscillators' amplitudes are equal $\alpha_1 = \alpha_2 = \alpha$ and the transmissivity $T$ of beamsplitters $BS_j$ is optimized to maximize the violation of the Bell inequality described further in the text.  Alice and Bob randomly choose between two possible measurement settings: $\{$\textit{on}, \textit{off}$\}$. For \textit{on} setting the local oscillators are switched on i.e. $\alpha \neq 0$. For \textit{off} setting local oscillator is turned off i.e. $\alpha=0$. Note that the transmissivity of the local beamsplitters $BS_j$ is fixed. The setting change requires only blocking the local oscillator, which could be easily realized, e. g. with shutters. This feature is an additional advantage from the point of view of experimental feasibility. 

After each run of the experiment, both parties assign binary  outcomes  $\{0,1\}$ related to the number of photons  $n_{c_j}$  and $n_{d_j}$ measured in their local detectors. The assignment depends on the chosen settings and goes in the following way:
$ 
\\
\text{for \textit{on} setting}:  
\left\{
	\begin{array}{ll}
		0: & \mbox{if at least one photon is detected in the mode $d_j$} \\ &\mbox{and no photons in mode $c_j$  \  i.e. \ } (n_{d_j}>0 \wedge n_{c_j}=0 ) 
		\\
		1: & \mbox{ for all other events } 
	\end{array}
\right.
\\
\text{for \textit{off} setting}:  
\left\{
	\begin{array}{ll}
		0:  & \mbox{if any photon is detected \ i.e. \ } (n_{d_j}>0 \vee n_{c_j}>0 )   \\
		1: & \mbox{for all other events } 
	\end{array}
\right.
$

Note that such an assignment is  perfectly tailored for the use of binary photodetectors. Let us denote conditional probability distributions in this setup as $P(ab|xy)$ in which $a$ and $b \in \{0,1\}$ and $x$ and $y \in \{$\textit{on}, \textit{off}$\}$. This distribution is given by the following:
\begin{eqnarray}
 P_{AB}(ab|xy)    = \begin{array}{cc|cc|cc}
			&\multicolumn{1}{c}{x} & \multicolumn{2}{c}{off}	& \multicolumn{2}{c}{on} \\
			y & \text{\diagbox[width=2.2em, height=2.2em, innerrightsep=0pt]{$b\,\,\,\,\,\,$}{$a\,\,\,$}}  & 0	& 1	&0  & 1 \\
			\hline \\ [-0.9em]
			\multirow{2}{*}{$off$ }  & 0 & 0	& \frac{1}{2}		& \xi_{00}& \xi_{01}\\[0.3em]
			& 1	& \frac{1}{2}	& 0	& \xi_{10}& \xi_{11} \\ [0.2em]
			\hline  \\ [-0.9em]
			\multirow{2}{*}{$on$ }   &0 &  \xi_{00} & \xi_{10} & \beta_{00}
   & \beta\\ [0.3em]
			& 1 & ~~\xi_{01}~~ & ~~\xi_{11}~~ & ~~\beta~~
   & ~~\beta_{11}~~
		\end{array},~~~~
		\label{box_positive}
\end{eqnarray}
where $\xi_q,\,\beta_q$ are functions of $\alpha$ and  $T$ and their exact forms are given in the Appendix \ref{app:prob}.

It is clearly  noticeable that  for (\textit{off}, \textit{off})  we observe perfect anti-correlations as only a single photon from the signal beam is present in the system and detection of the photon can occur only in one of the measurement stations. 
Thus,  the outcomes from runs  (\textit{off}, \textit{off}) will be used to generate a secret raw  key.   A part of the (\textit{off}, \textit{off}) outcomes will be used ( together with the results obtained from the three other possible  choices of settings) to perform Bell test. The test is done to check the security of the  distributed secret key. In order to ensure that the secret raw key is long enough, Alice and Bob should choose \textit{off} setting more often than  \textit{on}.  

When the experiment is done, Alice reveals to Bob all her settings and the part of results, which are needed to perform a Bell test or in general to perform a search for eavesdropper optimal strategy, i.e. all \textit{on} and the part of  \textit{off} results. These results combined with Bob's local ones allow Bob to estimate the key rate and to perform standard privacy amplification procedures. 
For the scheme described above, the following Clauser-Horne (CH) inequality will be used: 
\begin{equation}
   CH = P (00 | on, on)+P (00 | on, off)+P (00 | off, on)-P (00 | off, off)  
    - P_A (0 | on) - P_B (0 | on)  \leq 0.\label{eq:ch}
\end{equation}
where $P_A(a|x),\,P_B(b|y)$ denotes local probabilities.

\begin{figure}
     \centering
         \includegraphics[width=\columnwidth*2/3]{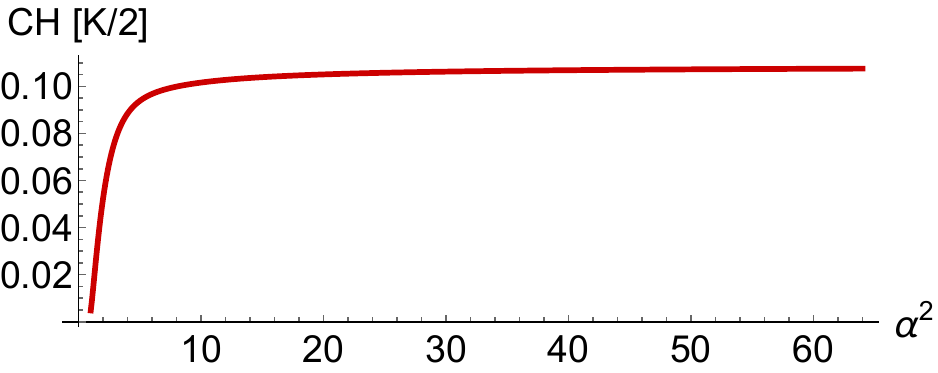}

        \caption{Plot of the optimal violation of the right hand side of the Clauser-Horne inequality, given in  (\ref{eq:ch}),  as a function of the  local oscillator strength $\alpha^2$, for
         state $\ket{\psi}_{b_1,b_2}$.
         The value of the CH expression is optimized  with respect to the beamsplitter transmissivity  $T$ for each value of $\alpha^2$ and it determines the value of the secret key rate by the  formula $K=2 CH$. }\label{fig:ch}
\end{figure}

Inequality (\ref{eq:ch}) can be violated  for some value of $\alpha$ given an optimized value of $T$.
FIG. \ref{fig:ch} shows how the maximal value of the CH expression in (\ref{eq:ch}) increases as a function of  the square of the amplitude $\alpha^2$. To assess maximal violation, we consider the limiting case of $\alpha\rightarrow\infty$, with the constraint that the product $\alpha^2T$ converges to a finite value, i.e. $\alpha^2T \rightarrow k$. Then the limiting value of the  $CH$  expression is given by:
\begin{equation}
  \lim_{\alpha^2  \rightarrow \infty}
\lim_{\alpha^2 T \rightarrow k}  CH = e^{-2 k} \left(2-e^k\right) k,\label{eq:ch_inf}
\end{equation}
and it reaches maximal value of $CH \approx 0.1086$  for $k \approx 0.260$. Looking at the Fig. \ref{fig:ch}  one notices that in practice the CH value saturates rapidly as a function of $\alpha^2$  and one is close to the maximal violation already for $\alpha^2 \approx 10$.

To sum up, we present short outline of the protocol:
\begin{enumerate}
    \item Alice and Bob agree on the choice of the local oscillators' amplitude $\alpha$ and with respect to it they set local beamsplitters' transmissivity $T$, so that the violation of CH inequality is possible for the setup. 
    \item Single photon pairs are generated in PDC process. When the detector placed in the idler mode clicks, Alice and Bob  start the measurement with one of the local settings chosen by each of them. The choice between the settings \textit{on} and \textit{off} is random and with no previous agreement between the parties. However, there should be sufficiently more \textit{off} than \textit{on} choices, as \textit{off} settings are used to test the security as well as to generate the  raw key. 
    \item The signal photon impinges on the balanced beamsplitter and the resulting output field is sent to the detection station of  Alice and Bob.
    \item At the detection stations local weak homodyne measurement is performed.   
    \item After sufficient number of experimental runs Alice communicates her settings and the part of her results needed for Bell test to Bob, who performs the Bell test (or in general performs search for eavesdropper optimal strategy) and privacy amplification. If inequality is not violated, Bob terminates the key sharing procedure.
    \item Raw key is obtained from a fraction of runs of the experiment in which local settings where set to  (\textit{off}, \textit{off}).
\end{enumerate}

\section{Security}

\subsection{Outline}

Our security proof  assumes that Eve has full control over the preparation of the physical system that mediates and imitates the correlations between Alice and Bob. The entire system is described in terms of joint no-signalling probability distributions.
This is the  most general approach to tackle the question of security, as any protocol which is secure against individual attacks by no-signaling Eve is also secure against \textit{any} individual attacks of Eve able to control quantum resources used by Alice and Bob. 

The clue of the proof is the mathematical property of no-signaling correlations, which is that whenever in a tripartite scenario a Bell inequality violation is observed between two parties (Alice and Bob), the third party (Eve), who governs the process of preparation of the correlations cannot have deterministic knowledge about \textit{all} the outcomes of Alice and Bob. The degree of Eve's uncertainty about Alice and Bob's outcomes can be directly expressed in terms of the degree of violation of a Bell inequality. In order to do it, we utilize the theory of decomposition of  no-signaling distributions into  mixtures  consisting of either fully deterministic correlations or maximally non-local (but still no-signaling) correlations, for which the local outcomes are totally random. By means of recent developments in this theory we find a distribution which for a given value of a Bell inequality violation observed by Alice and Bob ensures a minimal amount of admixture of non-local correlations in the decomposition. Such a strategy is an optimal one for Eve, who in this way minimizes the privacy of Alice and Bob's outcomes. Based on this minimal amount of privacy, we find a lower bound on the secret key rate generated between Alice and Bob  as a function of observed Bell inequality violation.

\subsection{The proof}

\begin{figure}
    \centering
    \includegraphics[width=0.5\linewidth]{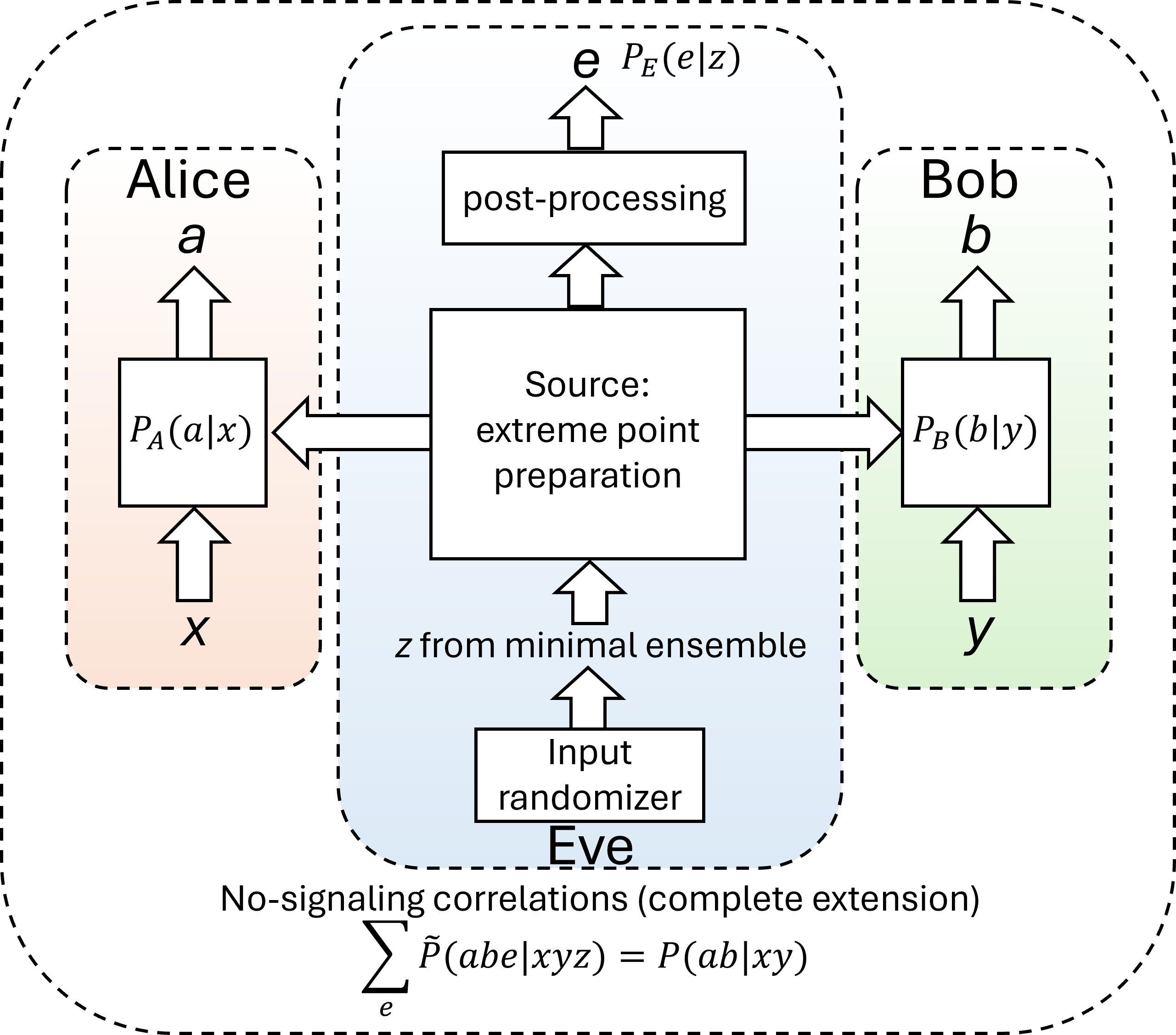}
    \caption{Schematic representation of eavesdropper preparing no-signaling correlations mimicking correlations of Alice and Bob. To this matter Eve prepares state to randomly correspond to one of the extreme points of no-signaling polytope. Then Eve keeps part of the system for herself and perform measurement on it to extract information about Alice's and Bob's subsystems (here Eve measures which extreme point she prepared). }
    \label{fig:eve}
\end{figure}

To prove security of our protocol let us introduce eavesdropper (Eve) into the system. We consider the presence of no-signaling Eve which has full control over the state preparation and performs individual attacks in each run of the protocol (see FIG. \ref{fig:eve}).

In each run, Eve prepares a tripartite
system from which two subsystems are distributed to
Alice and Bob, and the third subsystem is kept by her. Afterwards Eve can choose some measurement from set $Z \equiv\{z\}$  to perform on her subsystem, resulting in some outcome $e$ used by her to extract information. We assume that the probability distribution $\Tilde{P}(abe|xyz)$ of the extended system is no-signaling, i.e., it fulfills no-signaling conditions: 
\begin{equation}
\label{nsCond}
    \sum_e \Tilde{P}(abe|xyz) = P(ab|xy), ~~\forall z \in Z 
\end{equation} 
and analogous conditions for Alice and Bob. This corresponds to the statement that the no-signalling
Eve cannot influence the statistics of outcomes of the
honest parties through her choices of the measurement
settings. What is more, Eve is able to prepare the system in such a way that she can achieve any $\Tilde{P}(abe|xyz)$ that ideally mimics the quantum device assumed by Alice and Bob characterized by probability distribution $ P(ab|xy)$ to avoid exposing herself. Note that $P(ab|xy)$, as a quantum probability distribution, is by definition
 no-signaling between the parties. Such an eavesdropper can be restated  in terms of the theory of no-signaling correlations as Eve holding \textit{complete extension} \cite{CE} of quantum device $P(ab|xy) $. \textit{Complete extension} is a generalization of the purification of the quantum state into no-signaling correlation boxes. This gives ultimate power to no-signaling Eve for individual attacks in the same sense as using purification by quantum Eve is optimal. For more details on the theory of \textit{complete extensions} see Appendix \ref{app:complete_extension} or for a full discussion \cite{CE}.
 
\subsection{Secret key rate}
In order to calculate secret key rate we have to find an optimal strategy for Eve. Let us consider the strategy in which Eve decomposes the quantum probability distribution $P(ab|xy)$, treated now as a no-signaling correlation box,  into the extreme points of the no-signaling polytope corresponding to a scenario with two inputs and two outputs per party. By extreme points, we mean those that cannot be written as a convex combination of other points from the correlation polytope. It is known that this polytope contains $24$ extreme points among which $16$ are local boxes and  $8$ are non-local boxes. They are characterized by respectively four and three binary indices, and their formal definitions are provided in the Appendix \ref{app:complete_extension} (see formulas \eqref{local_boxes} and \eqref{nonlocal_boxes}). Here let us
 recall for example local box $L_{0011}$ and non-local box $B_{111}$:
\begin{eqnarray}
L_{0011}    = \begin{array}{cc|cc|cc}
			&\multicolumn{1}{c}{x} & \multicolumn{2}{c}{off}	& \multicolumn{2}{c}{on} \\
			y & \text{\diagbox[width=2.2em, height=2.2em, innerrightsep=0pt]{$b\,\,\,\,\,\,$}{$a\,\,\,$}}  & 0	& 1	&0  & 1 \\
			\hline \\ [-0.9em]
			\multirow{2}{*}{$off$ }  & 0 & 0	& 0		&0 & 0\\[0.3em]
			& 1	& 1	& 0	& 1& 0\\ [0.2em]
			\hline  \\ [-0.9em]
			\multirow{2}{*}{$on$ }   &0 &  1 & 0 & 1
   & 0\\ [0.3em]
			& 1 & ~~0~~ & ~~0~~ & ~~0~~
   & ~~0~~
		\end{array},~~~~\,\,\,\,\,\,\, 
		B_{111}    = \begin{array}{cc|cc|cc}
			&\multicolumn{1}{c}{x} & \multicolumn{2}{c}{off}	& \multicolumn{2}{c}{on} \\
			y & \text{\diagbox[width=2.2em, height=2.2em, innerrightsep=0pt]{$b\,\,\,\,\,\,$}{$a\,\,\,$}}  & 0	& 1	&0  & 1 \\
			\hline \\ [-0.9em]
			\multirow{2}{*}{$off$ }  & 0 & 0	& \frac{1}{2}		& \frac{1}{2}& 0\\[0.3em]
			& 1	& \frac{1}{2}	& 0	& 0& \frac{1}{2} \\ [0.2em]
			\hline  \\ [-0.9em]
			\multirow{2}{*}{$on$ }   &0 &  \frac{1}{2} & 0 & \frac{1}{2}
   & 0\\ [0.3em]
			& 1 & ~~0~~ & ~~\frac{1}{2}~~ & ~~0~
   & ~~\frac{1}{2}~~
		\end{array}.
		\label{eq:box_example}
\end{eqnarray}
Based on the Carath\'{e}odory theorem \cite{caratheodory,rockafellar1997convex} each of the element (box) in the polytope can be decomposed at most into $9$ extreme boxes \cite{CE}. The weights with which particular boxes appear in the decomposition can be simply interpreted as the probabilities that Eve prepares the subsystem of Alice and Bob in such a way that it exactly matches correlations of particular extreme box. Such decompositions are called \textit{minimal ensembles} \cite{CE} if one cannot find another decomposition using only a subset of extreme boxes used in a given decomposition.  Note that there is always finite number of such ensembles. Furthermore,  they can be easily and efficiently calculated by solving a set of linear equations (see Appendix \ref{app:minimal_ensemble} for details). Eve after preparing some \textit{minimal ensemble} performs measurement on her subsystem to reveal which no-signaling box was shared with Alice and Bob. If the shared box is local, Eve gains perfect deterministic knowledge about the results of the run. However, if the box is non-local, she has zero knowledge about results for key generation runs, as for this case they are completely random (see for example (\ref{eq:box_example})). 

The key point is that allowing Eve to prepare any \textit{minimal ensemble} enables her to perform any no-signaling strategy for individual attacks. This is because in this scenario Eve can effectively reach any decomposition of $P(ab|xy)$ in terms of any points in the no-signaling polytope by properly using the input randomizer (random choice of \textit{minimal ensemble}) and the output post-processing channel.
Note that, post-processing channel  due to the data-processing inequality  cannot improve mutual information between Bob and Eve or Alice and Eve \cite{Data_processing}. Therefore, no strategy using post-processing channels can be more optimal, and we can focus on input randomization. In Appendix \ref{app_en_box} we provide the set of all \textit{minimal ensembles} for $P(ab|xy)$  for
a pair of settings $\{\alpha, T\}$ that provide near optimal violation of CH inequality (\ref{eq:ch}), namely for $\alpha\rightarrow\infty$ and $T$ such that $\alpha^2T$ tends to some finite number $k$. We stress that we find analogous ensembles for all $\alpha$ considered in the Fig. \ref{fig:ch} with only slightly changed probabilities. Among those \textit{minimal ensembles} there is one, which we denote as $\mathbf{M_{op}}$ that has the smallest total probability of Eve  sending non-local boxes $p_{nl}$. Note that $p_{nl}\neq0$ is implied by the violation of CH inequality. As Eve wants to perform optimally she has to minimize the probability of non-local boxes as these do not give her any information. Therefore, Eve will always choose the optimal setting corresponding to ensemble $\mathbf{M_{op}}$ as otherwise she will introduce more randomness to her results, i.e. she would more often have to guess the bit as non-local boxes would appear with higher frequency. Note that, Eve's optimal strategy is independent of the choice of settings of Alice and Bob. Therefore, Eve cannot gain additional knowledge about particular bits from the information about the settings used after measurements of Alice and Bob are done.  Thus, this information can be shared between Alice and Bob without any concern after their measurements. 

Knowing optimal strategy we can calculate secret key rate for one way communication using Csisz\'ar-K\"orner formula \cite{Key_rate_one_way}:
\begin{equation}
    K=max\{I(A:B)-I(A:E),I(A:B)-I(B:E)\},\label{eq:key}
\end{equation} 
where $I(X:Y)$ stands for mutual information. Due to the perfect anti-correlation between Alice and Bob for key generation runs we have $I(A:B)=1$. Further, due to the symmetry of our setup, we have $I(B:E)=I(A:E)$. What is more, as Eve gets perfect knowledge from local boxes, mutual information between Bob and Eve is simply equal to the probability of sending local box by Eve $p_l=1-p_{nl}$. Finally, the key rate is given just as $K=p_{nl}$. In the Appendix \ref{app:CH_vs_K} section we show based on Eve optimal strategy that there is the following relation between the key rate and optimal CH inequality violation:
\begin{equation}
    K=2CH.\label{eq:CH_K}
\end{equation}

Based on FIG. \ref{fig:ch} we observe that with growing $\alpha$ the key rate grows to its optimal value $K\approx0.217$ for $\alpha\rightarrow\infty$ which can be found based on eq. (\ref{eq:ch_inf}). However, the maximal key rate is quickly saturated and already for $\alpha^2=8$ the key rate is $K\approx0.2$. Note that the optimal strategy for Eve, in principle, can vary if one introduces different noise structures, and therefore the relation (\ref{eq:CH_K}) is not necessarily general. However, in such a case, one can still use algorithm for finding the optimal strategy for Eve through \textit{minimal ensembles} to calculate the key rate for this specific correlation box. 

\section{Self-testing random number generator}
Our proposed experimental setup can also be used as a self-testing random number generator (RNG) \cite{rngRev}. In this context, the task of the protocol is to generate provably random bits, without the need for perfect correlation between bits possessed by Alice and Bob, as in the case of secret key generation. The RNG protocol itself is exactly the same as for key distribution, what is different is the role played by non-classical correlations and the security proof. In the case of QKD correlations shared between Alice and Bob play a double role: first for assuring perfect correlations between shared random bits, second for testing security of the protocol by measuring Bell inequality violation. In the case of RNG protocol, the only role of shared correlations between Alice and Bob is to test the quantumness of the optical state impinging on the central beam splitter, and therefore the true quantum nature of the randomness produced by the RNG protocol. This is again done by measuring the violation of the Bell inequality. 

Let us first present the general idea of device-independent security proof of RNG protocol (see \cite{Brunner13}, sec. IV.C.3). The main entity in the proof is the average probability of Eve guessing Alice's outcome (in this presentation we focus on Alice's device as the RNG source, whereas Bob's device serves just for obtaining Bell inequality violation; due to symmetry of our scenario, the roles of Alice and Bob can be exchanged without any modification). Let us assume, as in the previous case, that Alice, Bob and Eve share joint correlations $\tilde P(abe|xyz)$ prepared by Eve of any nature, with the only restriction that they are no-signalling. Then the maximal average guessing probability can be defined using reduced conditional probability distributions \cite{Brunner13}:
\begin{equation}
    \label{pGuess}
    p_{\textrm{guess}}=\max_z\sum_e P(e|z)\max_a P(a|e,z).
\end{equation}
We assume that Alice and Bob perform a Bell test, which leads to the experimental estimation of an expression $\mathcal B$ of some Bell inequality. In specific scenarios, in particular ours, guessing probability can be expressed as a function of the Bell value $\mathcal B$: $ p_{\textrm{guess}}= p_{\textrm{guess}}(\mathcal B)$.
The next step of the proof is introduction of the so-called \textit{min-entropy}, which measures the randomness of a discrete random variable in the strictest way among all entropies defined within the family of Renyi entropies \cite{Konig09}. The mean entropy is defined as the minus logarithm of the most probable outcome, which in the context of our problem translates to:
\begin{equation}
    \label{minEnt}
    H_{\textrm{min}}(\mathcal B)=-\log_2\left(p_{\textrm{guess}}(\mathcal B)\right).
\end{equation}
Min-entropy in our case can be treated as a function of the Bell value $\mathcal B$, which will be crucial in our argumentation.  $H_{\textrm{min}}$ well approximates the maximal number of secure random bits per run of the protocol that can be extracted from raw data in a given protocol using optimal privacy amplification techniques (known in this context as randomness extraction techniques) \cite{RennerPHD, Konig09}:
\begin{equation}
    \label{lbOnRandomBits}
   l_{\textrm{random}}\approx H_{\textrm{min}}.
\end{equation}
As was shown in fundamental works on quantum randomness generation \cite{Vazirani11, De12, Berta12}, this relation holds also in the quantum scenario. Namely, if all the correlations are due to sharing quantum states. Therefore, in device-independent schemes for quantum random number generation, the number of securely random bits in a given protocol involving the Bell inequality test is a function of the measured Bell value $\mathcal B$:
\begin{equation}
    l_{\textrm{random}}\approx -\log_2\left(p_{\textrm{guess}}(\mathcal B)\right).
\end{equation}

Let us now apply the above presented proof to the case of our protocol. We will utilize the super-quantum no-signalling description of correlations used in the previous sections. Although there are no known results on the precise relation between the number of secure bits and entropy in the case of such general model of correlations, the relation \eqref{lbOnRandomBits} calculated for the general no-signalling distribution gives the lower bound for the number of extractable random bits in the case of quantum correlations, since they are no-signalling. More precisely, the relation  \eqref{lbOnRandomBits} for no-signalling correlations gives the minimal number of extractable random bits for any quantum implementation of the ideal individual eavesdropping attack. To calculate this number, let us first calculate the guessing probability \eqref{pGuess}. In the model of Eve preparing arbitrary non-signalling distribution the guessing probability of Alice's (or Bob's) outcome does not depend on Alice's (or Bob's) settings. As in the case of key rate analysis we can assume without loss of generality that Eve prepares some \textit{minimal ensamble} of extreme boxes. Then there are just two possibilities for guessing probability. If Eve's outcome indicates that a local box has been prepared, the guessing probability equals $1$, as this case is fully deterministic, whereas if non-local box has been prepared, the guessing probability equals $\tfrac{1}{2}$, since the outputs are then totally random. Therefore, guessing probability for Eve equals:
\begin{equation}
\label{pg1}
    p_{\textrm{guess}}=p_l+\frac{1}{2}p_{nl}=1-\frac{1}{2}p_{nl},
\end{equation}
where the second equality comes just from normalisation $p_l+p_{nl}=1$. Now note that $p_{\textrm{guess}}$ \eqref{pg1} is maximized for minimal $p_{nl}$, therefore, it attains maximal value for the same optimal decomposition $\mathbf{M_{op}}$ as discussed in the case of optimal Eve's attack against secret key generation. We already know that for this decomposition, the probability $p_{nl}$
is equal to twice the $CH$ value: $p_{nl}=2CH$, therefore we finally have:
\begin{equation}
\label{pg2}
    p_{\textrm{guess}}=1-CH.
\end{equation}
This gives the approximate number of truly random bits that can be extracted when using our protocol:
\begin{equation}
   l_{\textrm{random}}\approx -\log_2\left(1-CH)\right).
\end{equation}
For the maximal violation $CH\approx 0.1086$ we obtain $ l_{\textrm{random}}\approx 0.1660$. 
Note that violation of the Bell inequality and thus also presence of non-local boxes in the $\mathbf{M_{op}}$ ensures that the origin of the random bits can be attributed to the fundamental quantum randomness emerging from nonexistence of local hidden variable models.

\section{Discussion}

In summary, the setup called by Tan, Walls and Collett ``nonlocality of a single photon'' can be used, after necessary modifications, to design a device-independent cryptographic scheme secure against even a ``non-signaling'' Eve. Our protocol links some features of the pioneering BB84 and of (a generalized) Ekert91 protocol:  the secret key is distributed using a single photon interacting with a beamsplitter and the security is based upon observed Bell non-classicality due to entanglement. However, this nonclassicality is induced by the single-photon superposition which is equivalent to an entangled state of two quantum optical modes. The entanglement does not involve pairs of spatially separated photons. 

The advantage of the protocol is that it requires the minimal possible number of pairs of settings, that is just four, for performing a Bell test,  and  a specific distinguishing feature of it is that it  uses just one pair of these settings, the \textit{off} ones, in the process of key generation whereas the previous protocols had to add an additional setting for the key generation. This allows for the security analysis in terms of the simplest non-signaling polytope and thus the least computationally demanding security check against no-signaling eavesdropping. This can be important for smooth and fast operations of potential commercial devices based on such solutions. 

The security analysis is based on a search for the optimal strategy for Eve through the decomposition of obtained correlations to extreme points of the no-signaling polytope. The method used is independent of the construction of the device and could also be applied in different scenarios, for example, in the analysis of the proposed setup in the presence of noise. We also apply this method to propose the scheme for quantum random number generation. The obtained probability of guessing the bit by the third parties allows us to compute lower bounds for the bit rate. This shows the versatility of the analysis used. 

It is important to note that while the security analysis method based on the correlations decomposition is  device independent, a relation between the violation of a specific CH inequality with the key rate and bit rate obtained by the decomposition of correlations relies on the specific setup (device) considered, and these relations do not have to be general. This is because the relations for the violation of CH inequality were calculated for a specific class of correlations present in the perfect implementation of the proposed experimental setup, and while it could hold in the noisy scenario it is not guaranteed. The presence of noise in correlations could result in a different decomposition of the correlations being optimal for Eve, potentially resulting in different relations of key rate and bit rate with the used CH inequality. Thus, to obtain unconditional security and accurately calculate key rate or bit rate one should apply the whole decomposition procedure, but still one can treat CH inequality violation as a quick estimate if the security was not compromised beforehand. This is because obtaining a violation of some Bell inequality is a necessary condition for obtaining a nonzero key rate and bit rate. The violation of CH inequality can also be used for device calibration, as the higher violation of CH inequality indicates a more significant presence of not favorable for Eve non-local correlations, which results in higher key rate and bit rate. Still, further developments of the protocol in terms of noisy systems may provide a general proof of the relation between the violation of the CH inequality and the key rate.

The proposed experimental setup should be possible to implement in the laboratory with current technology, as it requires only standard quantum optical equipment and high-efficiency threshold single-photon detectors, which are commercially available. Furthermore, the measurement itself in its principles is a homodyne measurement, which has been used in laboratories for decades with the difference that used local oscillator fields are weak. Still, experiments involving such techniques were successfully conducted in recent years \cite{WALMSLEY}.

Finally, note that arbitrary no-signalling correlations are impossible to observe or engineer. Still they allow to construct simple criteria for secure transmission by modeling on eavesdropper constrained only by the rules of special relativity. 
Therefore, even though our method is highly general and computationally not demanding, one could opt to consider only quantum strategies for eavesdropping. Because a set of quantum correlations is a subset of a set of no-signaling correlations, such an approach results in at least as high key rate as for no-signaling Eve and therefore could provide faster key distribution without real compromises on security. 
However, analyzing of all possible quantum strategies is a demanding problem worthy of further investigation, even in this simplest possible set of configurations. One can also try to provide security only in some subclasses of quantum eavesdropping strategies specific to the device that are most probable to be implemented; however, this is a non-trivial research problem as the considered system is infinite-dimensional.

\section*{Acknowledgements}

BW and TD thank the ‘International Centre for
Theory of Quantum Technologies’ project (contract no. 2018/MAB/5), which is carried out
within the International Research Agendas Programme (IRAP) of the Foundation for Polish
Science (FNP) co-financed by the European Union from the funds of the Smart Growth
Operational Programme, axis IV: Increasing the research potential (Measure 4.3). 

\noindent KS, MM, and MŻ thank the IRA Programme, project no. FENG.02.01-IP.05-0006/23, financed by the FENG program 2021-2027, Priority FENG.02, Measure FENG.02.01., with the support of the FNP.


\section*{Appendices}
\appendix
\section{Probabilities for all versus nothing events}\label{app:prob}
In this section, we show how probabilities    $P(ab|xy)$ can be calculated. To simplify the problem, let us restate  measurement setting \textit{off}  for both the parties in such a way that it corresponds to the removal of the $i$-th party local beamsplitter $BS_i$  (which is equivalent to the situation of a beamsplitter of $100\%$ transmissivity) and outcomes $0$ and $1$ are assigned as in the case of \textit{on} setting. One can easily check that such a scenario is exactly equivalent to the one described in the main text.
We consider the following state:
\begin{equation}
    \ket{\psi(\alpha,\phi)}=\frac{1}{\sqrt{2}}\ket{\alpha e^{i\phi}}_{a_1}\left(\ket{01}_{b_1,b_2}+\ket{10}_{b_1,b_2}\right)\ket{\alpha e^{i\phi}}_{a_2},
\end{equation}
where for \textit{off} setting $\alpha=0$.  The probability of measuring a specific combination of photons in modes $d_i,c_i$ is given by:
\begin{equation}
    P(k,l,n,m|x,y)=|_{c_1,d_1,c_2,d_2}\bra{k,l,n,m}\hat U_1(x)\hat U_2(y)\ket{\psi(\alpha,\phi)}|^2,
\end{equation}
where  $\hat U_i(x)$ denotes unitary operator corresponding to the  beamsplitter $BS_i$ acting on the $i$-th party for setting $x$. This unitary performs the following transformation of annihilation operators for \textit{on} setting:
\begin{equation}\label{beamspilt}
\left( \begin{array}{c}
\hat c_j \\
\hat d_j 
\end{array} \right) =
\left( \begin{array}{cc}
\cos{\chi} & i\sin{\chi} \\
i\sin{\chi} &\cos{\chi}
\end{array} \right)
\left( \begin{array}{c}
\hat a_{j} \\
\hat b_{j} 
\end{array} \right) 
\end{equation}
where $\chi = \cos^{-1}(\sqrt{T})$. For \textit{off} setting we have $\hat U_i(off)=\mathbb{I}$. Based on that, probabilities $P^{\alpha,T}_{AB}(ab|xy)$ can be calculated as:
\begin{align}
        &P(00|x,y)=\sum_{l=0}^\infty\sum_{m=0}^{\infty}P(0,l,0,m|x,y),\\
         &P(10|x,y)=\sum_{k=1}^\infty\sum_{l=0}^{\infty}\sum_{m=0}^{\infty}P(k,l,0,m|x,y)+\sum_{m=0}^{\infty}P(0,0,0,m|x,y),\\
          &P(01|x,y)=\sum_{l=0}^\infty\sum_{n=1}^{\infty}\sum_{m=0}^{\infty}P(0,l,n,m|x,y)+\sum_{l=0}^{\infty}P(0,l,0,0|x,y),\\
          &P(11|x,y)=1-P(00|x,y)-P(10|x,y)-P(01|x,y).
\end{align}
The explicit expressions for probabilities $P_{AB}(ab|xy)$ with $\chi = \cos^{-1}(\sqrt{T})$ are given by:
\begin{align}
 \begin{split}
 P(00|on,on)& = \beta_{00}=\frac{1}{8} (2 \cos (2 \chi )+2) e^{\alpha ^2 (-\cos (2 \chi ))-\alpha ^2} \\&\times\left(-2 \alpha ^2 \cos (2 \chi
   )+e^{-\alpha ^2 \sin ^2(\chi )} \left(\alpha ^2 \cos (2 \chi )-\alpha ^2-2\right)+2 \alpha ^2+2\right),\end{split}\\
    \begin{split}
      P(11|on,on)& = \beta_{11}=\frac{1}{8} \Big(4 \cos ^2(\chi ) e^{-3 \alpha ^2 (\cos (2 \chi )+2)} \\&\left(e^{\frac{1}{2} \alpha ^2 (5 \cos (2 \chi
   )+9)} \left(\alpha ^2 \cos (2 \chi )-\alpha ^2-2\right)-4 e^{\alpha ^2 (2 \cos (2 \chi )+5)} \left(\alpha ^2 \cos
   (2 \chi )-\alpha ^2-1\right)\right)\\
   &+e^{-2 \alpha ^2 \cos ^2(\chi )} \left(8 e^{2 \alpha ^2 \cos ^2(\chi
   )}+e^{\alpha ^2 \cos ^2(\chi )} \left(\alpha ^2 \cos (4 \chi )-\alpha ^2-4 \cos (2 \chi )-12\right)\right.\\&\left.+2 \alpha ^2
   \cos (4 \chi )+8 e^{\alpha ^2 \cos (2 \chi )}-2 \alpha ^2-4 \cos (2 \chi )-4\right)\Big),\end{split}
   \end{align}
   \begin{align}
   P(10|on,on)&=P(01|on,on)=\beta= \frac{1}{2}\Big(1-P(11|on,on)-P(00|on,on)\Big) = \frac{1}{2}(1 - \beta_{00} + \beta_{11}) \\
     P(00|on,off)&= P(00|off,on) = \xi_{00}=\frac{1}{2} e^{-\alpha ^2} \left(e^{-\frac{1}{4} \alpha ^2 (2 \cos (2 \chi )-2)}-1\right),\label{eq:p00OnOff}\\
      P(11|on,off)&= P(11|off,on) = \xi_{11}=\frac{1}{2} \left(1-\cos ^2(\chi ) \left(\alpha ^2 \sin ^2(\chi )+1\right) e^{-\alpha ^2 \cos ^2(\chi )}\right),\\
   P(10|on,off)&=P(01|off,on) = \xi_{01}=\frac{1}{2} \left(e^{-\alpha ^2}-e^{-\alpha ^2 \cos ^2(\chi)}+1\right),\label{eq:p10OnOff}\\
    P(01|on,off)&=P(10|off,on) = \xi_{10} = \frac{1}{2} \cos ^2(\chi ) \left(\alpha ^2 \sin ^2(\chi )+1\right) e^{-\alpha ^2 \cos ^2(\chi )},
\end{align}

\section{Coherence (entanglement) swapping }\label{app:swaping}
Quantum repeaters \cite{repeaters} seem to be one of the necessary tools to perform QKD at long distances. The process on which quantum repeaters rely is entanglement swapping which is well known for singlet states \cite{Ent-swapp}. However, the analogous process called coherence swapping \cite{coh-swapp} for two copies of single-photon state (\ref{input-state}) is not widely known. Therefore, let us recall this basic idea, which would allow for the construction of the quantum repeaters tailored for the proposed QKD algorithm. Consider four modes $b_1,b_2$ and $b_1',b_2'$ with both pairs of modes in state (\ref{input-state}). Suppose that one wants to entangle the modes $b_1$ and $b_2'$. To achieve this, a symmetric beam-splitting operation (\ref{beamspilt}) is performed on the modes $b_2$ and $b_1'$, which are transformed to the output modes $c_1,c_2$. Then the resulting state has the following form:
\begin{equation}
    \frac{1}{2}\left( \hat b_1^\dagger \hat b_2'^\dagger+\hat c_1\frac{1}{\sqrt{2}}\big[\hat b_2'^\dagger-i\hat b_1^\dagger\big]+\hat c_2\frac{1}{\sqrt{2}}\big[\hat b_1^\dagger-i\hat b_2'^\dagger\big]-\frac{i}{2}\big[(\hat c_1^\dagger)^2+(\hat c_2^\dagger)^2\big]\right)\ket{\Omega}.
\end{equation}
Clearly if one detects a single photon across modes $c_1$ and $c_2$, then the state of the modes $b_1$ and $b_2'$ can be transformed by local unitary operation to the form (\ref{input-state}). Thus one performs coherence swapping in this setup with success rate $\tfrac{1}{2}$. \ks{Let us also note, that if one does not register coincidence of idler photons in both   }  \mrk{heralding detectors, then simply the coherence swapping procedure is not initiated, equivalently, the run is invalid.}

\section{Short review of theory of Complete extensions}\label{app:complete_extension}
Here we present short review of theory of complete extensions.

\textbf{Non-signalling device:}
A conditional probability distribution $P_{AB}(ab|xy)$, shared between two parties Alice and Bob, where $x$ and $a$ are the input and output choices of Alice and $y$ and $b$ are the input and outputs of Bob, will be called a non-signalling device if it is positive
\begin{equation}\label{device_positive}
    0 \leq P_{AB}(ab|xy) \leq 1, ~~\forall a,b,x,y
\end{equation}
satisfied the normalization condition
\begin{equation}\label{device_normalization}
    \sum_{ab} P_{AB}(ab|xy) = 1, \forall x,y
\end{equation}
and the non-signalling conditions
 \begin{eqnarray}\label{device_nonsig}
 \sum_a P_{AB}(ab|x,y)  = P_B(b|y), ~~\forall b,x,y \\
 \sum_b P_{AB}(ab|x,y) =  P_A(a|x). ~~\forall a,b,y \label{device_nonsig2}
 \end{eqnarray}
Here the cardinality of the set of all inputs and outputs are finite.

\textbf{No-signalling Polytope:} The entire state-space of all sets of non-signalling devices $P_{AB}(ab|xy)$ of same cardinalities  of inputs $x,y$ and  outputs $a,b$,   is called no-signalling polytope.  In general it is a convex set,  subset of $\mathbb{R}^N$, for an integer $N$, and  bounded by the linear constraints \eqref{device_positive}, \eqref{device_normalization}, and \eqref{device_nonsig} and \eqref{device_nonsig2}.

\textbf{Ensemble of $P_{AB}(ab|xy)$:} Let us for simplicity from now on omit the set of input and outputs $(ab|xy)$, from the notation of a device. An arbitrary non-signalling device $P_{AB}$ 
lies in a polytope can always be expanded as a convex combination of some of the  other members $\{Q^i\}_{i = 1}^n$ of the polytope, such that 
\begin{equation}
    P_{AB} = \sum_{i=1}^n q_i Q^i,
\end{equation}
holds for $q_i > 0, ~~\forall i$, and $\sum_i^n q_i = 1 $. Then the set of ordered pair $\{(q_i,  Q^i)\}$, will be called an ensemble of $P_{AB}$. Note that there can be infinitely many ensembles of a given device, and $q_i$ can be called as the probabilities of getting device $Q^i$.

\textbf{extreme (Pure) points:} A  no-signalling device $\texttt{E}$, will be called an extreme point or a pure point of that polytope if it can not expanded in terms of the other devices of the polytope.
As any polytope has to satisfy some set of linear constraints, hence every polytope (bounded polyhedron), has some finite number of pure points.

\textbf{Pure members ensembles \cite{CE}:} An ensemble of $P_{AB}$, will be called a pure members ensemble (PME) if all the member devices  are pure (extreme) points in the polytope. The ensemble $\{(p_i,  \texttt{E}^i)\}$ is a PME, as $P_{AB} = \sum_i p_i \texttt{E}^i$, and $\sum_i p_i = 1$, and the members are pure.
Note that the decomposition $\{p_i\}$, need not be unique.  

\textbf{Minimal ensembles \cite{CE}:} A  pure members ensemble $\{(p_i,  \texttt{E}^i)\}_{i \in \mathcal{I}}$, of $P_{AB}$, will be called a minimal ensemble, if any proper subset $U$ of the set of members, $U \subset \{\texttt{E}^i\}_{i \in \mathcal{I}}$, along with new choices of convex combination $\{\tilde{p}_j\}$, cannot form any PME of $P_{AB}$. Where $\mathcal{I}$ is any index set. The probabilities $\{p_i\}$, of any minimal ensemble is \textit{unique}.

Consider  one exemplary PME of $P_{AB}$, which contains only $4$ pure members, 
say
\begin{equation}PME_0 = \{(p_1,  \texttt{E}^1), (p_4,  \texttt{E}^4), (p_9,  \texttt{E}^9), (p_{13},  \texttt{E}^{13}) \},\end{equation}
where $p_1 + p_4 + p_9 + p_{13} = 1$, and the index set is $\mathcal{I}_0 = \{1,4,9,13\}$. 
 Now if $PME_0$, is a minimal ensemble of $P_{AB}$, then there exists no subset 
$U \subset \{\texttt{E}^1, \texttt{E}^4, \texttt{E}^9, \texttt{E}^{13}\}$, with a new set of convex combination $\{\tilde{p}_j\}$, which is also a PME of $P_{AB}$, i.e., $P_{AB} \neq \sum_{j , \texttt{E}^j \in U} \tilde{q}_j \texttt{E}^j$. 

With the help of all the above definition, we are now going to write the definition of \textit{complete extension}, ${\cal E}(P)_{ABE}(abe|xyz)$, of the bipartite device $P_{AB}(ab|xy)$, where the subsystem $E$, has been controlled by the non-signalling eavesdropper, with $z$ being her input and $e$ is her output. The \textit{complete extension}, of a given device gives the no-signalling eavesdropper the ultimate operational power \cite{CE}, which the  quantum purification provides to a quantum Eve, for quantum device dependent \cite{Ekert1991,Brunner13,Mayers-Yao,acin-2007-98, Masanes2011, lit12} and quantum device independent scenarios \cite{Kent,AcinGM-bellqkd, masanes-2006, acin-2006-8, hanggi-2009}. 

\textbf{Complete extension \cite{CE}:} Given a bipartite device $P_{AB}(ab|xy)$, an extension ${\cal E}(P)_{ABE}(abe|xyz)$, will be called the \textit{complete extension} of the given device, \textit{iff} for any $z = k$, and $e = j$, we have 
\begin{equation}
    {\cal E}(P)_{ABE}(abe|xyz) = p(e = j|z = k) P_{AB}^{jk}(ab|xy),
\end{equation}
such that $\{(p(e = j|z = k), P_{AB}^{jk}(ab|xy))\}$, is a minimal ensemble of $P_{AB}(ab|xy)$, and corresponding to each minimal ensemble of $P_{AB}(ab|xy)$, there exist one input $z = k$, which generates it. Here,by $j,k$ in $P_{AB}^{jk}(ab|xy)$, we mean the register which Eve will possess to keep the track of which extreme box created in part of Alice and Bob. 

Suppose that $P_{AB}$, is a binary input output device and for this device, we know that there are only $24$ extreme points $\texttt{E}^i$ among them $16$ are local deterministic one and $8$ are completely non-local devices \cite{Barrett_extremal}. 
They can be characterised in the following way: 16 local boxes:
\begin{equation}\label{local_boxes}
    \mathrm{L}_{qrst}(ab|xy) = \left\{ \begin{array}{cc}
         1 & \mbox{if $a= q x\oplus r$,} \\
         	  & ~~~\mbox{$b = s y \oplus t$} \\
        0 & \mbox{otherwise}.\end{array} \right.
\end{equation}
with $q,r,s,t \in \{0,1\}$ and $x,y=0$ for \textit{off} setting and $x,y=1$ for \textit{on} setting. Remaining  $8$ non-local boxes can be put as: 
\begin{equation}\label{nonlocal_boxes}
    \mathrm{B}_{rst}(ab|xy) = \left\{ \begin{array}{ll}
         1/2 & \mbox{if $a\oplus b = xy\oplus rx \oplus sy \oplus t$} \\
        0 & \mbox{otherwise}.\end{array} \right.
\end{equation}

In Appendix \ref{app_en_box}, we enumerate all possible minimal ensembles of the correlation box for our experimental scheme for $\alpha\rightarrow\infty$. One of such minimal ensembles is 
\begin{eqnarray}
\mathbf{M_1}=\{(p_1,L_{0100}),      (p_1,L_{1110}),   (p_2,L_{1100})  ,  (p_3,L_{1001}),     (p_4,B_{101}),     (p_5,B_{011}) ,      (p_6,B_{111})\},    
\end{eqnarray}
Here the subscript $1$ in $M_1$, indicates that the measurement choice of Eve is $z = 1$, and after this choice of measurement Eve can create the ensemble $M_1$, in part of Alice and Bob. And with probability $p(e = 1|z = 1) = p_1$, the conditional box in part of Alice and Bob is $ P_{AB}^{j = 1, k = 1}(ab|xy) = L_{0100}$, and with probability $p(e = 2|z = 1) = p_1$, the conditional box in part of Alice and Bob is $ P_{AB}^{j = 2, k = 1}(ab|xy) = L_{1110}$ and so on, where $\sum_e p(e|z = 1) = 1$.

For each no-signalling box, there exists only a finite number of minimal ensembles. And the complete extension which is the no-signaling analogue of quantum purification, is not extreme device in the higher dimensional state space but it satisfies two crucial properties of quantum purification, namely ACCESS and GENERATION, which are the most important properties to consider the secret key agreement protocol.

\textbf{ACCESS:} A complete extension, ${\cal E}(P)_{ABE}(abe|xyz)$, of a device $P_{AB}(ab|xy)$, together with access to arbitrary randomness, in part of the extending system, gives access to any ensemble of the given device.

\textbf{GENERATION:} A complete extension, ${\cal E}(P)_{ABE}(abe|xyz)$, of a device $P_{AB}(ab|xy)$, with the access of input randomizer and output post-processing channel can be transformed to any no-signalling extension $Q_{ABE}(abe'|xyz')$.

\section{Finding minimal ensembles}\label{app:minimal_ensemble}
Here we present how one can find a minimal ensemble for some no-signalling correlation box for two-input-two-output scenario, but this method can be easily generalized to other scenarios.
For such a case there are 24 extreme points in the no-signalling polytope \cite{Barrett_extremal}.

In order to find all minimal ensembles one can treat the correlation box as a matrix which we denote as $\mathbb{P}(ab|xy)$. Now, let us choose some subset $\mathcal{I}_s$ of the set of  indices $\mathcal{I}=\{1,...,24\}$ with the power of this subset $\overline{\mathcal{I}_s}\leq 9$. The reason for choosing the maximal power of the set is that  each point in the regarded polytope  can be decomposed at most using $9$ extreme points. The assumed decomposition can be written using those extreme boxes $\texttt{E}^i$ in the form of a matrix from the set of all extreme boxes $\{\texttt{E}^i\}$ with $i\in\mathcal{I}_s$:
\begin{equation}
    \mathbb{P}(ab|xy)=\sum_{\mathcal{I}_s}w_i\texttt{E}^i,
\end{equation}
where $w_i$ are weights of the decomposition. This matrix equation is simply over-determined set of $16$ linear equations for values of each matrix element with $\overline{\mathcal{I}_s}$ variables. One can find such coefficients $w_i$ using pseudo-inverse  that will minimize the overall square error in the set of equations. However, whenever decomposition exists the error is equal to $0$. To check if coefficients provide proper decomposition one has to check the following criteria:
\begin{align}
    \begin{split}
      ||    \mathbb{P}(ab|xy)-\sum_{i\in\mathcal{I}_s}w_i\texttt{E}^i||_1<\epsilon,
    \end{split}\\
    \begin{split}
        |1-\sum_{i\in\mathcal{I}_s}w_i|<\epsilon,
    \end{split}\\
    \begin{split}
        \forall_{i\in\mathcal{I}_s} \;\;w_i>0,
    \end{split}
\end{align}
where $\epsilon$ stands for small arbitrarily picked number which is introduced due to the possible small numerical errors during calculations. These criteria ensure that the coefficients obtained provide the proper convex decomposition of  $\mathbb{P}(ab|xy)$. If criteria are fulfilled, one takes the decomposition as a candidate for minimal ensemble. To obtain all minimal ensembles, one has to perform this procedure for all possible subsets $\mathcal{I}_s$. In the last step, one has to check if all candidates are linearly independent and eliminate those that can be put as a convex combination of other candidates. Note that this step is not necessary when looking for optimal strategy for Eve, as any convex combination of minimal ensembles cannot have a lower probability of non-local boxes than the lowest probability among those particular minimal ensembles. Therefore Bob during security check does not need to eliminate non-minimal ensembles.

Note that this method allows for decomposing box that is slightly deviating from the non-signaling polytope. This is important for practical use, as measured frequencies that estimate probabilities do not have to be perfectly no-signalling.

\section{Relation with CH inequality violation}\label{app:CH_vs_K}
Here we consider the relation between CH inequality violation and the key rate. Let us consider the optimal decomposition of a correlation box  (\ref{box_positive}) for the optimal $T$. It has the following form for  $\alpha$ from the range presented on Fig. 2 in the main text and also for $\alpha\rightarrow\infty$:
\begin{eqnarray}
\mathbf{M_{opt}}=\{(q_1,L_{0011}), (q_1,L_{1100}),    (q_2,L_{0110}),     (q_2,L_{1001}),     (q_3,L_{1011}),     (q_3,L_{1110}),     (p_{nl},B_{111})\}, \label{eq:optimal_dec2}
\end{eqnarray}
where $q_i$ stands for weights of the decomposition. This decomposition can be also put directly in a form of the probability box:
 \begin{equation}
\begin{array}{cc|cc|cc}
			&\multicolumn{1}{c}{x} & \multicolumn{2}{c}{\text{Off}}	& \multicolumn{2}{c}{\text{On}} \\
			y & \text{\diagbox[width=2.2em, height=2.2em, innerrightsep=0pt]{$b$}{$a~$}}  & 0	& 1	&0  & 1 \\
			\hline \\ [-0.9em]
			\multirow{2}{*}{Off }  & 0 & 0	& \frac{1}{2}		& \frac{1}{2}p_{nl}+q_1+q_3 & q_2\\[0.3em]
			& 1	& \frac{1}{2}	& 0	&  q_1 & \frac{1}{2}p_{nl}+q_2+q_3 \\ [0.2em]
			\hline  \\ [-0.9em]
			\multirow{2}{*}{On }   &0 &  \frac{1}{2}p_{nl}+q_1+q_3 & q_1 & \frac{1}{2}p_{nl}+2q_3 &  q_3 \\ [0.3em]
			& 1 & q_2 & \frac{1}{2}p_{nl}+q_2+q_3 & q_3&  \frac{1}{2}p_{nl}+2q_2
		\end{array}
  \label{eq:box_er}
\end{equation}

One can observe that the probability that Eve sends one of the local boxes $2(q_1+q_2+q_3)$, which is equal to the mutual information $I(B:E)=I(A:E)$ as discussed in previous sections,   can be found by adding elements of the box (\ref{eq:box_er}) for which corresponding elements of the box $B_{111}$ have value $0$ i.e diagonal elements for settings \textit{off, off} and anti-diagonal for other combinations of settings. Therefore we get:
\begin{equation}
I(B:E)=I(A:E)=2\left[P(10|on,on)-P(10|on,off)-P(01|on,off)\right],
\end{equation}
where we used the fact that:
\begin{equation}
    P(10|on,on)=P(01|on,on).
\end{equation}

From this and  from the fact that $I(A:B)=1$ for key generation, we can estimate the key rate using Csisz\'ar-K\"orner formula:
\begin{multline}
        K=max\{I(A:B)-I(A:E),I(A:B)-I(B:E)\}\\=1-2P(10|on,on)-2P(10|on,off)-2P(01|on,off).\label{eq:R_B}
\end{multline}

From the other side, let us consider expression of CH inequality (2) from the main text:
\begin{equation}
   CH = P (00 | on, on)+P (00 | on, off)+P (00 | off, on)  -P (00 | off, off)  
    - P_A (0 | on) - P_B (0 | on) . 
\end{equation}
 By noting that:
 \begin{align}
   & P(00|off,off)=0,\\
   & P(00|on,off)=P(00|off,on),\\
    & P_A(0|on)=P (01 | on, off)+P (00 | on, off),\\
    & P_B(0|on)=P (10 | on, on)+P (00 | on, on),
 \end{align}
the expression of CH inequality can be put as:
\begin{equation}
   CH= P (00 | on, off)-P (01 | on, off)-P (10 | on, on).\label{eq:CH_p_2}
\end{equation}
Subtracting two times CH expression (\ref{eq:CH_p_2}) from the key rate $K$ (\ref{eq:R_B}) we get:
\begin{equation}
    K-2CH=1-2(P (00 | on, off)+P (10 | on, off))=0,
\end{equation}
where we used (\ref{eq:p00OnOff}) and (\ref{eq:p10OnOff}) to get:
\begin{equation}
 P (00 | on, off)   +P (10 | on, off)=1/2.
\end{equation}
Thus, based on our assumptions, we get a relation of the key rate with CH inequality violation $K=2 CH$.

Note that we found the same structure of the optimal (from Eve's perspective) box decomposition (\ref{eq:optimal_dec2})  for all $\alpha$ considered in Fig. 2. Because the box $B(\alpha,\chi)$ is constructed from continuous functions of the settings parameters, the small deviations in settings will not significantly change the structure of the box. What is more, for increasing $\alpha$ deviations of the box decrease as the results converge to the results for $\alpha\rightarrow\infty$. This strongly suggests that relation $K=2 CH$ is true for any $\alpha$  with near optimal transmissivity for violation of CH inequality.

\section{Minimal ensemble for $\alpha\rightarrow\infty$}\label{app_en_box}
Let us consider as en example set of all minimal ensembles for $\alpha\rightarrow\infty$. For this case we have following correlation box:

\begin{multline}
\lim_{\alpha^2  \rightarrow \infty}
\lim_{\alpha^2 T \rightarrow k}   P_{AB}(ab|xy) =\\
  \begin{array}{cc|cc|cc}
			&\multicolumn{1}{c}{x} & \multicolumn{2}{c}{\text{Off}}	& \multicolumn{2}{c}{\text{On}} \\
			y & \text{\diagbox[width=2.2em, height=2.2em, innerrightsep=0pt]{$b\,\,\,\,\,\,$}{$a\,\,\,$}}  & 0	& 1	&0  & 1 \\
			\hline \\ [-0.9em]
			\multirow{2}{*}{Off }  & 0 & 0	& \frac{1}{2}		& \frac{e^{-k}}{2} &  \frac 12 - \frac{ e^{-k}}{2}\\[0.3em]
			& 1	& \frac{1}{2}	& 0	& \frac{k e^{-k}}{2}  & \frac 12 - \frac{ke^{-k}}{2} \\ [0.2em]
			\hline  \\ [-0.9em]
			\multirow{2}{*}{On }   &0 &   \frac{e^{-k}}{2} &  \frac{ke^{-k}}{2} &  2 e^{-2 k} k & \frac{e^{-2 k}}{2}  \left(e^k (k+1)-4 k\right) \\ [0.3em]
			& 1 &~~\frac 12 - \frac{e^{-k}}{2}~~ & ~~~\frac 12 - \frac{ke^{-k}}{2}~~~ & \frac{e^{-2 k}}{2}  \left(e^k (k+1)-4 k\right) &  2 e^{-2 k} k-e^{-k} (k+1)+1
		\end{array}
  \label{eq:box_infinity}
\end{multline}

We write all the minimal ensembles for the box $\lim_{\alpha^2  \rightarrow \infty}
\lim_{\alpha^2 T \rightarrow k}   P_{AB}(ab|xy)$ bellow as a sets of ordered pairs $\{(w_i,\texttt{E}^i)\}_{i\in\mathcal{I}_s}$.

\begin{eqnarray*}
\mathbf{M_1}=\{(p_1,L_{0100}),      (p_1,L_{1110}),   (p_2,L_{1100})  ,     (p_3,L_{1001}),     (p_4,B_{101}),     (p_5,B_{011}) ,      (p_6,B_{111})\},     \\
\mathbf{M_2}=\{(p_1,L_{0001}),     (p_1,L_{1011}),    (p_2,L_{0011}),      (p_3,L_{0110}),        (p_4,B_{011}) ,    (p_5,B_{101}),     (p_6,B_{111})\},     
\end{eqnarray*}

\begin{eqnarray*}
\mathbf{M_3}=\{(p_2,L_{0011}),     (p_7,L_{0100}),     (p_3,L_{1001}),     (p_7,L_{1110}),     (p_4,B_{101}),   (p_8,B_{011}),       (p_6,B_{111})\},     \\
\mathbf{M_4}=\{(p_2,L_{1100}),    (p_7,L_{0001}),     (p_3,L_{0110}),     (p_7,L_{1011}),      (p_4,B_{011}),     (p_8,B_{101}),     (p_6,B_{111})\},     
\end{eqnarray*}

\begin{eqnarray*}
\mathbf{M_5}=\{(p_1,L_{0110}) ,    (p_9,L_{1011}),   (p_9,L_{1110}),     (p_{10},L_{1100}),  (p_{11},L_{1001}),    (p_{5},B_{011}),     (p_{12},B_{111})\},    \\ 
\mathbf{M_6}=\{(p_{1},L_{1001}),     (p_{9},L_{1011})  ,   (p_{9},L_{1110}),    (p_{10},L_{0011}),     (p_{11},L_{0110}),    (p_{5},B_{101}),     (p_{12},B_{111})\},     
\end{eqnarray*}
\begin{eqnarray*}
\mathbf{M_7}=\{(p_{10},L_{0001})    , (p_{11},L_{0110}),     (p_{1},L_{1001}),     (p_{13},L_{1011}),     (p_{10},L_{1100}),     (p_{9},L_{1110}),     (p_{12},B_{111})\},     \\
\mathbf{M_8}=\{(p_{10},L_{0011}),     (p_{10},L_{0100}),     (p_{1},L_{0110}),     (p_{11},L_{1001}),     (p_{9},L_{1011}),     (p_{13},L_{1110}),     (p_{12},B_{111})\},     
\end{eqnarray*}

\begin{eqnarray*}
\mathbf{M_9}=\{(p_{1},L_{0100}),     (p_{14},L_{1001}),     (p_{1},L_{1110}),     (p_{2},B_{001}),     (p_{13},B_{011}),     (p_{15},B_{101}),     (p_{16},B_{111})\},     \\
\mathbf{M_{10}}=\{(p_{1},L_{0001}),     (p_{14},L_{0110}),     (p_{1},L_{1011}),     (p_{2},B_{001}),     (p_{15},B_{011}),     (p_{13},B_{101}),     (p_{16},B_{111})\},     
\end{eqnarray*}

\begin{eqnarray*}
\mathbf{M_{11}}=\{(p_{1},L_{0100}),     (p_{14},L_{1001}),     (p_{15},L_{1011}),     (p_{17},L_{1110}),     (p_{18},B_{001}),     (p_{14},B_{011}),     (p_{19},B_{111})\},     \\
\mathbf{M_{12}}=\{(p_{1},L_{0001}),     (p_{14},L_{0110}),    (p_{15},L_{1110}),    (p_{17},L_{1011}),       (p_{18},B_{001}),     (p_{14},B_{101}),     (p_{19},B_{111})\},     
\end{eqnarray*}
\begin{eqnarray*}
\mathbf{M_{13}}=\{(p_{1},L_{0100}),      (p_{20},L_{1011}),      (p_{9},L_{1110}),      (p_{21},L_{1100}),  (p_{11},L_{1001}),      (p_{5},B_{011}),     (p_{19},B_{111})\},     \\
\mathbf{M_{14}}=\{(p_{1},L_{0001}),      (p_{20},L_{1110}),      (p_{9},L_{1011}),      (p_{21},L_{0011}),  (p_{11},L_{0110}),      (p_{5},B_{101}),     (p_{19},B_{111})\}, 
\end{eqnarray*}
\begin{eqnarray*}
\mathbf{M_{15}}=\{(p_{20},L_{1011}),      (p_{21},L_{0011}),     (p_{10},L_{0100}),     (p_{11},L_{1001}),         (p_{17},L_{1110}),     (p_{14},B_{011}),     (p_{19},B_{111})\},     \\
\mathbf{M_{16}}=\{(p_{20},L_{1110}),     (p_{21},L_{1100}),      (p_{10},L_{0001}),     (p_{11},L_{0110}),     (p_{17},L_{1011}),     (p_{14},B_{101}),     (p_{19},B_{111})\},     
\end{eqnarray*}
\begin{eqnarray*}
\mathbf{M_{17}}=\{(p_{2},L_{0001}),     (p_{7},L_{0100}),     (p_{14},L_{1001}),     (p_{1},L_{1110}),     (p_{8},B_{011}),     (p_{4},B_{101}),     (p_{22},B_{111})\},    \\ 
\mathbf{M_{18}}=\{(p_{7},L_{0001}),     (p_{2},L_{0100}),     (p_{14},L_{0110}),     (p_{1},L_{1011}),     (p_{4},B_{011}),     (p_{8},B_{101}),     (p_{22},B_{111})\},     
\end{eqnarray*}
\begin{eqnarray*}
\mathbf{M_{19}}=\{(p_{2},L_{0001}),     (p_{2},L_{0100}),     (p_{1},L_{0110}),     (p_{1},L_{1001}),     (p_{8},B_{011}),     (p_{8},B_{101}),     (p_{22},B_{111})\},    
\end{eqnarray*}
\begin{eqnarray*}
\mathbf{M_{20}}=\{(p_{21},L_{0001}),     (p_{10},L_{0100}),     (p_{14},L_{1001}),     (p_{20},L_{1011}),     (p_{9},L_{1110}),     (p_{14},B_{011}),     (p_{22},B_{111})\},     \\
\mathbf{M_{21}}=\{(p_{21},L_{0100}),     (p_{10},L_{0001}),     (p_{14},L_{0110}),     (p_{20},L_{1110}),     (p_{9},L_{1011}),     (p_{14},B_{101}),     (p_{22},B_{111})\},     
\end{eqnarray*}

\begin{eqnarray*}
\mathbf{M_{22}}=\{(p_{10},L_{0001}),     (p_{10},L_{0100}),     (p_{1},L_{0110}),     (p_{1},L_{1001}),     (p_{9},L_{1011}),     (p_{9},L_{1110}),     (p_{22},B_{111})\},     
\end{eqnarray*}

\begin{eqnarray*}
\mathbf{M_{23}}=\{(p_{23},L_{0110}),     (p_{23},L_{1001}),     (p_{2},B_{001}),     (p_{13},B_{011}),     (p_{13},B_{101}),     (p_{16},B_{111})\},     
\end{eqnarray*}

\begin{eqnarray*}
\mathbf{M_{24}}=\{(p_{23},L_{0110}),     (p_{23},L_{1001}),     (p_{13},L_{1011}),     (p_{13},L_{1110}),     (p_{5},B_{001}),     (p_{12},B_{111})\},   
\end{eqnarray*}

\begin{eqnarray*}
\mathbf{M_{25}}=\{(p_{23},L_{0110}),     (p_{7},L_{1001}),     (p_{2},L_{1100}),     (p_{5},B_{011}),     (p_{8},B_{101})     (p_{6},B_{111})\},  \\ 
\mathbf{M_{26}}=\{(p_{23},L_{0110}),     (p_{7},L_{1001}),     (p_{2},L_{1100}),     (p_{5},B_{101}),     (p_{8},B_{011}),     (p_{6},B_{111})\},\\
\end{eqnarray*}

\begin{eqnarray}
\mathbf{M_{op}}=\mathbf{M_{27}}=\{(p_{10},L_{0011}),     (p_{11},L_{0110}),     (p_{11},L_{1001}),     (p_{13},L_{1011}),     (p_{10},L_{1100}),     (p_{13},L_{1110}),     (p_{24},B_{111})\}. 
\end{eqnarray}

\begin{align}
    \begin{split}
        p_1&=\frac{1}{2} e^{-k} \left(-k+e^k-1\right),\,\,
        p_2=\frac{1}{2} e^{-2 k} \left(e^k k+4 k-e^k\right),\,\,
        p_3=\frac{1}{2} e^{-2 k} \left(-e^k k+4 k-3 e^k+2 e^{2 k}\right),\\
         p_4&=-4 e^{-2 k} k+e^{- k} ( k+2)-1,\,\,
        p_5= e^{-k} k,\,\,
         p_6=e^{-k} (1-k),\,\,
        p_7=\frac{1}{2} e^{-2 k} \left(4 k-2 e^k+e^{2 k}\right),\\
        p_8&=e^{-2 k} \left(e^k-4 k\right),\,\,
        p_9=\frac{1}{2} e^{-2 k} \left(e^k-4 k\right),\,\,
        p_{10}=\frac{e^{-k} k}{2},\,\,
        p_{11}=\frac{1}{2} e^{-k} \left(e^k-1\right),\\
        p_{12}&=e^{-2 k} \left(4-e^k\right),\,\,
        p_{13}=\frac{1}{2} e^{-2 k} \left(e^k k-4 k+e^k\right),\,\,
        p_{14}=e^{-k} \left(-k+e^k-1\right),\\
        p_{15}&=\frac{1}{2} e^{-2 k} \left(-3 e^k k+4 k-3 e^k+2 e^{2 k}\right),\,\,
        p_{16}=\frac{1}{2} e^{-2 k} \left(e^k-e^k k+4 k\right),\,\,
        p_{17}=\frac{1}{2} e^{-2 k} \left(2-e^k\right) \left(e^k-2 k\right),\\
        p_{18}&=e^{-k} \left(2 k-e^k+1\right),\,\,
        p_{19}=e^{-2 k} \left(-2 e^k k+4 k-e^k+e^{2 k}\right),\,\,
        p_{20}=\frac{1}{2} e^{-2 k} \left(e^k k-4 k+2 e^k-e^{2 k}\right),\\
        p_{21}&=\frac{1}{2} e^{-k} \left(2 k-e^k+1\right),\,\,
        p_{22}=4 e^{-2 k} k,\,\,
        p_{23}=\frac{1}{2}  \left(1-(1+k)e^{-k}\right),\,\,
        p_{24}=2 e^{-2 k} \left(2-e^k\right) k.\\
    \end{split}
\end{align}

\bibliographystyle{quantum}
\bibliography{biblio}

\end{document}